\documentclass[twocolumn,showpacs,preprintnumbers,amsmath,amssymb,prl]{revtex4}

\usepackage{graphicx}
\usepackage{dcolumn}
\usepackage{bm}

\usepackage{mathptmx, courier, pifont}
\usepackage[scaled=0.92]{helvet}
\usepackage[T1]{fontenc}
\usepackage{textcomp}
\usepackage{color}
\usepackage{colordvi}

\newcommand{\quarterthin}{\kern 0.0417em}

\begin{document}


\title{Variation in displacement energies due to isospin nonconserving forces}

\author{K.~Kaneko$^{1}$, Y.~Sun$^{2,3}$, T.~Mizusaki$^{4}$, S.~Tazaki$^{5}$ }

\affiliation{
$^{1}$Department of Physics, Kyushu Sangyo University, Fukuoka
813-8503, Japan \\
$^{2}$Department of Physics and Astronomy, Shanghai Jiao Tong
University, Shanghai 200240, People's Republic of China \\
$^{3}$Institute of Modern Physics, Chinese Academy of Sciences,
Lanzhou 730000, People's Republic of China \\
$^{4}$Institute of Natural Sciences, Senshu University, Tokyo
101-8425, Japan \\
$^{5}$Department of Applied Physics, Fukuoka University, Fukuoka
814-0180, Japan
}

\date{\today}

\begin{abstract}
For mirror nuclei with masses $A=42-95$, the effects of isospin
nonconserving nuclear forces are studied with nuclear shell model
using the Coulomb displacement energy and triplet displacement
energy as probes. It is shown that the characteristic behavior of
the displacement energies can be well reproduced if the isovector
and isotensor nuclear interactions with $J=0$ and $T=1$ are
introduced into the $f_{7/2}$ shell. These forces, with their
strengths being found consistent with the nucleon-nucleon scattering
data, tend to modify nuclear binding energies near the $N=Z$ line.
At present, no evidence is found that these forces are needed for
the upper $fp$-shell. Theoretical one- and two-proton separation
energies are predicted accordingly, and locations of the proton
drip-line are thereby suggested.
\end{abstract}


\maketitle


Isospin is a fundamental concept in particle and nuclear physics
\cite{Wigner37}. Isospin-symmetry breaking occurs in particle
physics because of the $u$-$d$ quark mass difference and the
electromagnetic effects in quarks \cite{Miller06}. In nuclear
physics, nucleon-nucleon scattering data suggest that the
neutron-neutron ($nn$) interaction is $\sim$1\% more attractive than
the proton-proton ($pp$) interaction and the proton-neutron ($pn$)
interaction is $\sim$2.5\% stronger than the average of the $nn$ and
$pp$ interactions \cite{Henley69,Machleidt01a,Ormand89}. In nuclei,
the Coulomb interaction between protons also breaks both charge
symmetry and charge independence. The Coulomb displacement energy
(CDE), i.e. the binding-energy difference between mirror nuclei, is
a well-known signature of charge-symmetry breaking due to the
Coulomb interaction \cite{Bentley07}. However, it was realized
\cite{Nolen69} that even if the pairing, exchange, and
electromagnetic spin-orbit terms are considered, the Coulomb force
alone cannot account for the experimental CDE (known as the
Nolen-Schiffer anomaly). There have been many attempts to resolve
this discrepancy \cite{Shahnas94}. Shell-model calculations
suggested that the isospin nonconserving (INC) nuclear interactions
are important for understanding the anomaly \cite{Zuker02}. In
addition, one could also study the triplet displacement energy (TDE)
\cite{Garrett01}, which is regarded as a measure of breaking in
charge independence \cite{Bentley07}.

The study of proton-rich nuclei is one of the frontiers in
low-energy nuclear physics. Proton-rich nuclei with masses $A\sim
60-70$ are of particular interest. In this mass region, there are at
least three so-called waiting-points along the suggested path of
rapid proton capture process (the $rp$-process) \cite{Schatz06}:
$^{64}$Ge, $^{68}$Se, and $^{72}$Kr, having equal numbers of
neutrons and protons ($N=Z$). Precise masses in the vicinity of the
waiting-point nuclei
\cite{Blank95,Schury07,Savory09,Tu11,Rogers11,Robertson90} are
required to locate the $rp$-process path and to understand
astronomical observations on the abundance of chemical elements. The
concept of CDE is thought to be a reliable method for predictions of
unknown masses (or nuclear binding energies) on the proton-rich side
of the $N=Z$ line \cite{Ormand97,Brown02}. In addition, $N\sim Z$
nuclei with $A\sim80$ are known \cite{Janas99} to undergo dramatic
changes in shape \cite{Chandler97,Lister82} with addition or removal
of just one or two nucleons \cite{Lister90}, which would strongly
influence the determination of the end point of the $rp$-process
\cite{Schatz01}, i.e. the heaviest element that the $rp$-process
nucleosynthesis may create.

\begin{figure*}[t]
 \begin{center}
\includegraphics[totalheight=6.2cm]{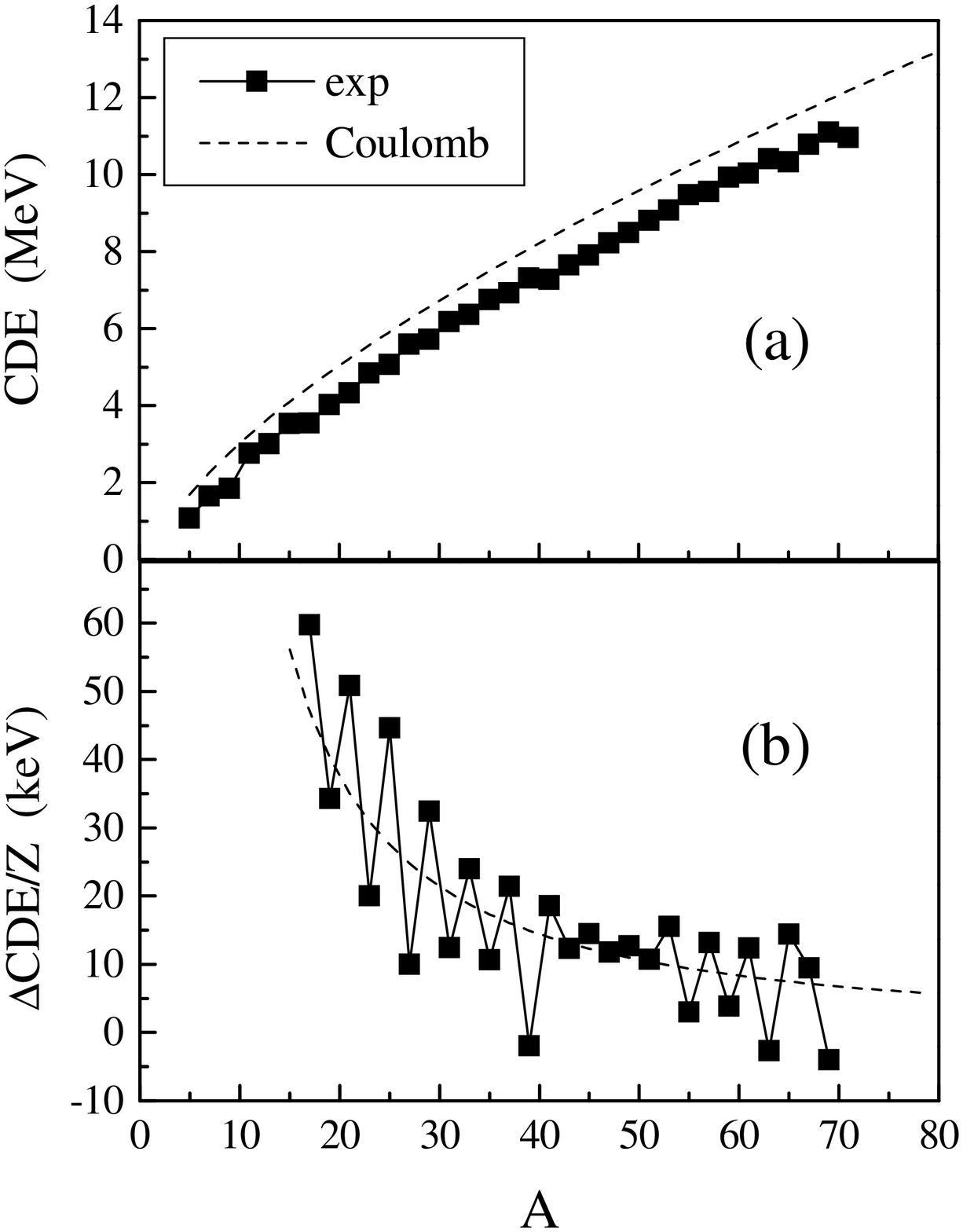}
\includegraphics[totalheight=6.2cm]{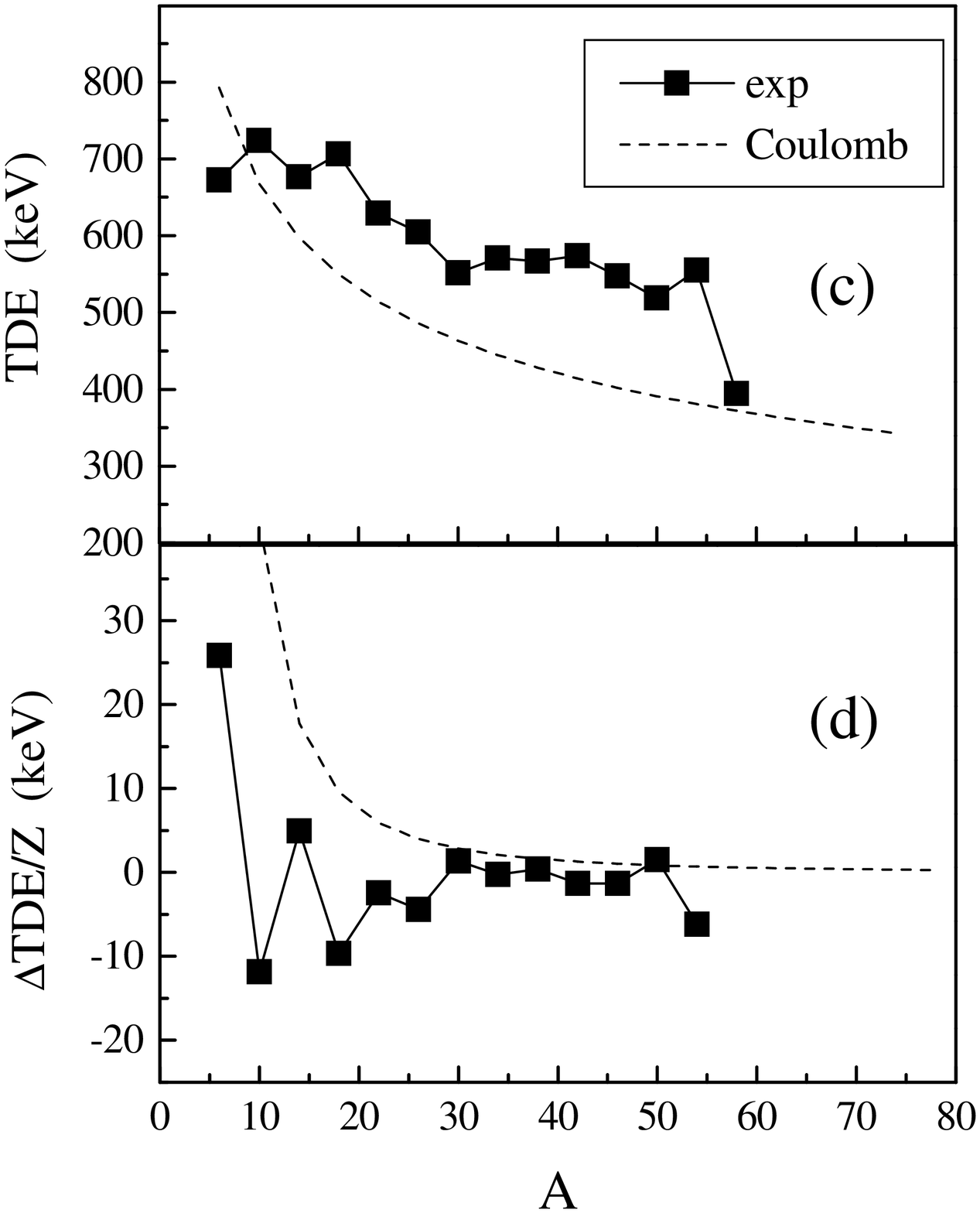}
  \caption{Experimental Coulomb displacement energy and triplet displacement energy.
(a) CDE, (b) differences in CDE between $A$ and $A+2$ nuclei (shown
as $\Delta$CDE/$Z$), (c) TDE, and (d) differences in TDE between $A$
and $A+4$ nuclei (shown as $\Delta$TDE/$Z$). Experimental data are
taken from Ref. \cite{Tu11,Audi03}. Theoretical curves from the
Coulomb prediction \cite{Bentley07} are shown for comparison.}
  \label{fig1}
\end{center}
\end{figure*}

The CDE for mirror nuclei is defined as
\begin{equation}
 {\rm CDE}(A,T) = BE(T,{T_{z}}_{<}) - BE(T,{T_{z}}_{>}),
  \label{eq:1}
\end{equation}
where $T_{z}=(N-Z)/2$ is the $z$ component of the total isospin $T$,
and $BE(T,{T_{z}}_{<})$ and $BE(T,{T_{z}}_{>})$ are (negative)
binding energies in an isospin multiplet having the largest proton
number ($Z_>$) and the smallest one ($Z_<$), respectively. For
$T=1/2$, the experimental CDE's \cite{Tu11,Audi03} are shown in Fig.
\ref{fig1}(a) and compared with the Coulomb energy prediction
\cite{Bentley07}. A monotonous increasing trend in CDE with
increasing mass number is described for the entire region from $A=5$
to 71. However, an overall overestimate by the calculation is seen
in Fig. \ref{fig1}(a). These deviations from data can be
qualitatively understood by the exchange effects due to the Pauli
Principle, which keeps the protons apart, thus weakening the Coulomb
repulsion \cite{Nolen69,Bentley07}. A close examination on the curve
indicates a zigzag behavior in these CDE's. To see the zigzag
pattern more clearly, we introduce a quantity measuring the
differences in CDE between nuclei $A$ and $A+2$,
\begin{equation}
 \Delta{\rm CDE}(A,T) = {\rm CDE}(A+2,T) - {\rm CDE}(A,T).
\label{eq:2}
\end{equation}
In Fig. \ref{fig1}(b), one clearly sees an odd-even staggering
pattern. The Coulomb energy prediction gives only the average with a
smooth curve. A notable exception in the pattern is seen for the
$f_{7/2}$-shell nuclei with masses $A=42-52$, where the staggering
seems to be washed out considerably.

The TDE with $T=1$ is defined with binding energies of triplet
nuclei as
\begin{equation}
 {\rm TDE}(A,T) = BE(T,{T_{z}}_{<}) + BE(T,{T_{z}}_{>}) - 2BE(T,T_{z}=0).
  \label{eq:4}
\end{equation}
In Fig. \ref{fig1}(c), the known experimental TDE's \cite{Audi03}
are shown for different masses. Except for those around $A=$ 6 and
at $A=$ 58, the Coulomb energy prediction disagrees strongly with
data, particularly for those $f_{7/2}$-shell nuclei where much
enhanced TDE's are observed experimentally. We may further introduce
a quantity measuring the differences in TDE,
\begin{equation}
 \Delta{\rm TDE}(A,T) = {\rm TDE}(A+4,T) - {\rm TDE}(A,T).
  \label{eq:5}
\end{equation}
In Fig. \ref{fig1}(d), it is seen that staggering occurs only for light
nuclei, but fades away for heavier ones. For nuclei starting from $A=$ 30,
the experimental $\Delta$TDE's show a smooth behavior, with only $A=$ 54
as an exception.

Questions arise as to why in the mass region of $A=42-54$ the
staggering magnitude in $\Delta$CDE is greatly reduced, why the
overall TDE is significantly larger than the Coulomb prediction for
this mass region, and why the $\Delta$TDE at $A=54$ (and TDE at
$A=58$) suddenly deviates from the smooth trend. To find an answer,
we perform state-of-the-art shell-model calculations with inclusion
of the INC interaction $H_{INC}$ in addition to the original
isoscalar Hamiltonian $H_{0}$. For $H_{0}$, we adopt two modern
interactions: GXPF1A \cite{Honma05} with the full $fp$ shell and
JUN45 \cite{Honma09} with the $pf_{5/2}g_{9/2}$ model space. The
total Hamiltonian then reads
\begin{equation}
 H = H_{0} + H_{INC},
  \label{eq:6}
\end{equation}
where $H_{INC}$ takes the form of a spherical tensor of rank two
\begin{equation}
 H_{INC} = H'_{sp} + V_C + \sum_{k=1}^{2}V_{INC}^{(k)},
  \label{eq:7}
\end{equation}
with $V_C$ in Eq. (\ref{eq:7}) being the Coulomb interaction and
$H'_{sp}$ the single-particle Hamiltonian that includes the Coulomb
single-particle energy for protons and the single-particle energy
shifts $\varepsilon_{ls}$ due to the electromagnetic spin-orbit
interaction for both protons and neutrons with the parameters taken
from Ref. \cite{Andersson05}. The Coulomb single-particle energies
for protons are taken as (all in MeV) $\varepsilon(0f_{7/2})=7.4$,
$\varepsilon(1p_{3/2})=7.2$, $\varepsilon(0f_{5/2})=7.1$, and
$\varepsilon(1p_{1/2})=7.3$ for the $fp$ model space, and
$\varepsilon(1p_{3/2})=9.4$, $\varepsilon(1f_{5/2})=9.1$,
$\varepsilon(1p_{1/2})=10.0$, and $\varepsilon(0g_{9/2})=9.7$ for
the $pf_{5/2}g_{9/2}$ model space. The electromagnetic spin-orbit
term has been shown to play an important role for understanding the
anomalies in the Coulomb energy difference in $^{67}$As/$^{67}$Se
\cite{Kaneko10} and $^{70}$Br/$^{70}$Se \cite{Kaneko12}. The
$\varepsilon_{ll}$ term \cite{Duflo02} does not appear explicitly
because this term shifts only the proton single-particle energies
and are effectively included in the Coulomb single-particle energies
listed above. $V_{INC}^{(k)}$ in (\ref{eq:7}) is the INC
interaction, with $k=1$ and $k=2$ for the isovector and isotensor
component, respectively. The two-body matrix elements with $T=1$ are
related to those in the proton-neutron formalism
\cite{Ormand97,Bentley07} through
\begin{equation}
 V_{INC}^{(1)}  =  V_{pp} - V_{nn}, ~~~~
 V_{INC}^{(2)}  =  V_{pp} + V_{nn} - 2V_{pn},
  \label{eq:8}
\end{equation}
where $V_{pp}$, $V_{nn}$, and $V_{pn}$ are, respectively, the $pp$,
$nn$, and $pn$ matrix elements of $T=1$.

Calculations are performed for odd-mass nuclei with isospin $T=$
1/2, 3/2, and 5/2 and for even-mass nuclei with $T=$ 1, 2, and 3, in
both the $fp$ and $pf_{5/2}g_{9/2}$ model spaces. Because of large
dimensions involved in the calculation, it is necessary to restrict
the number of nucleons to be excited from the lower to the upper
orbits. We have carefully checked the results between calculations
with and without restrictions and found that they differ by only a
few keV.

\begin{figure*}[t]
 \begin{center}
\includegraphics[totalheight=6.4cm]{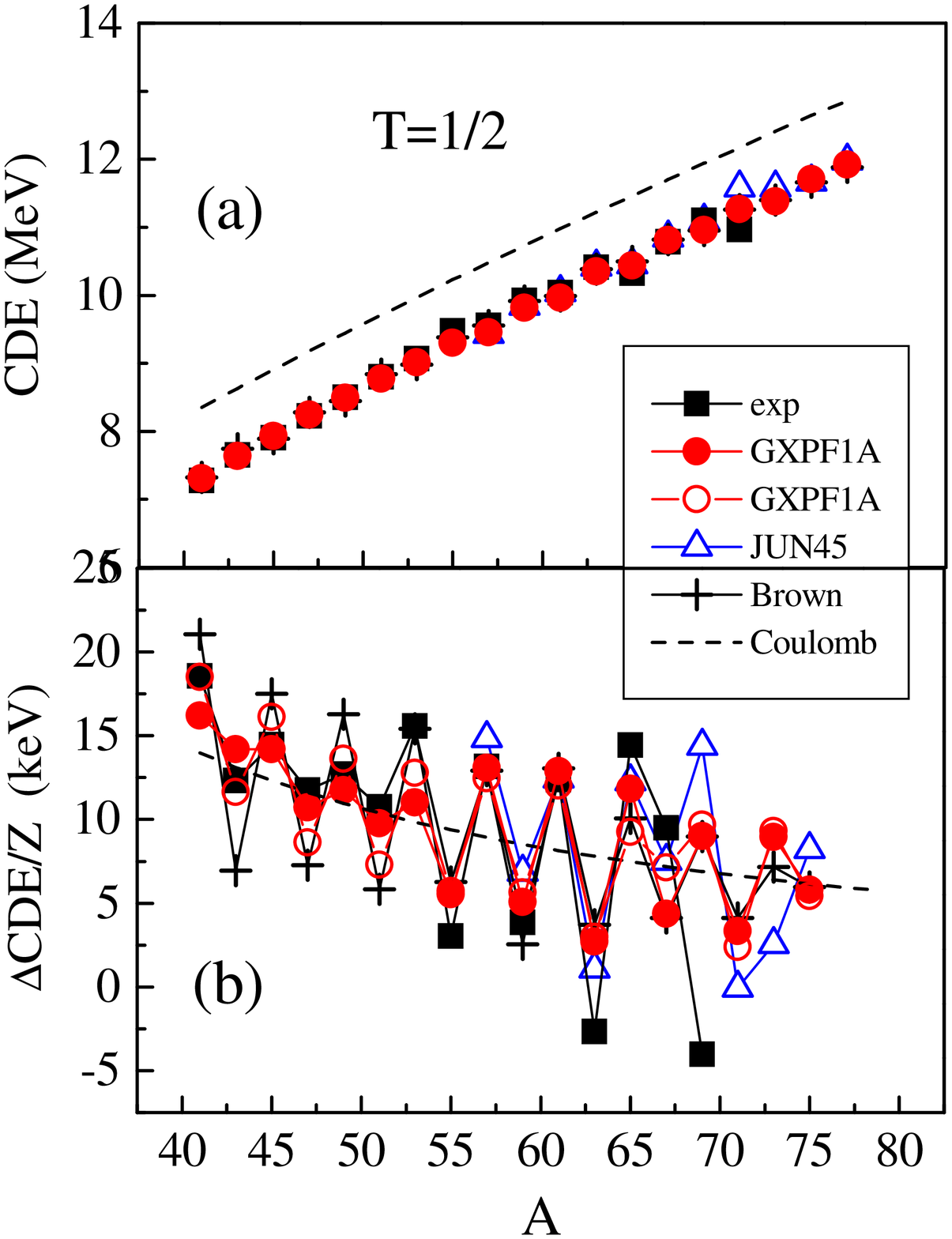}
\includegraphics[totalheight=6.4cm]{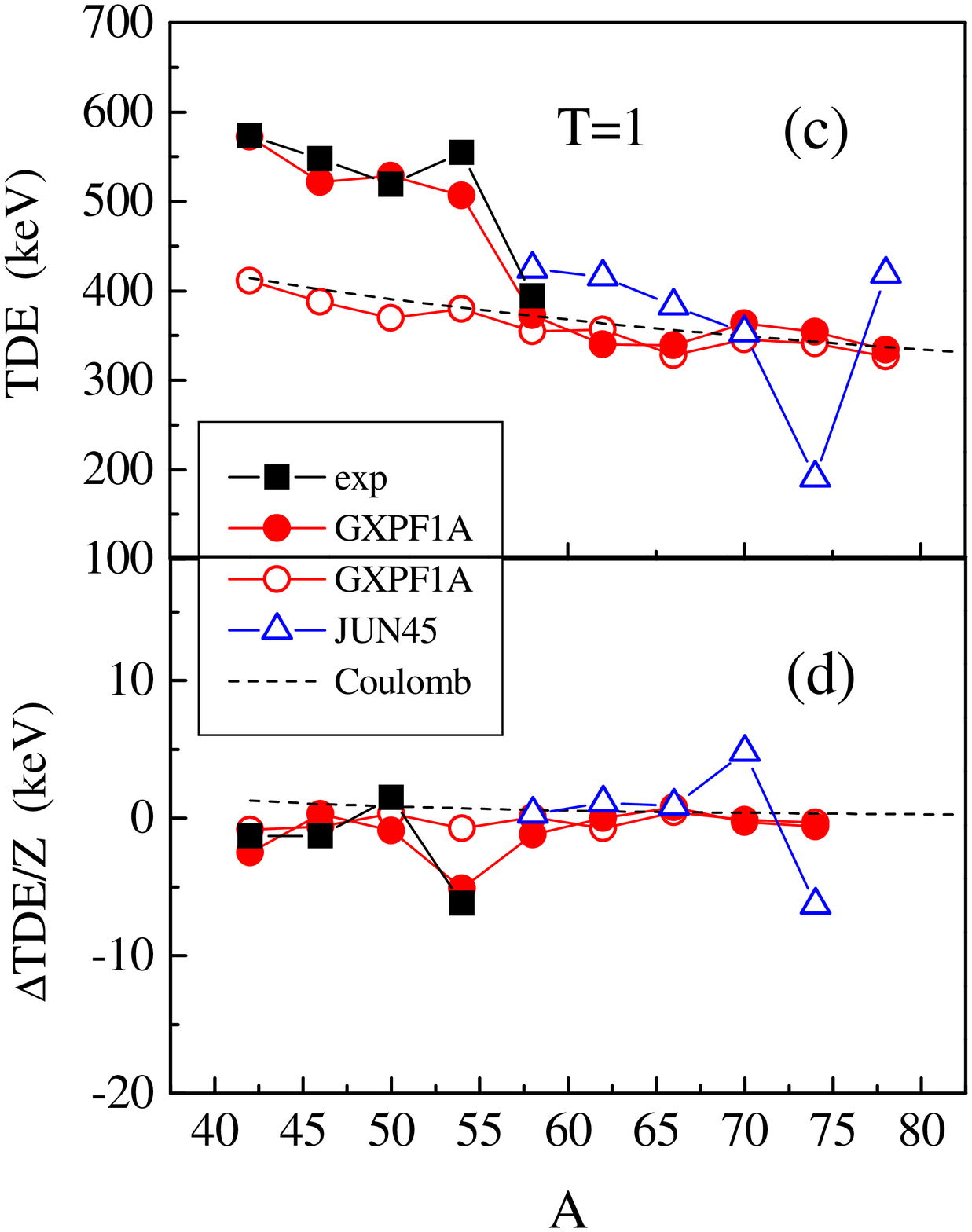}
\caption{(Color online) Calculations with GXPF1A and JUN45
  interactions are
  compared with experimental data \cite{Tu11,Audi03} for (a) CDE, (b)
  ${\Delta\rm CDE}(A,T)/Z$, (c) TDE, and (d) ${\Delta\rm TDE}(A,T)/Z$.
  Solid (open) symbols indicate results with (without) the INC nuclear
  interactions in the $f_{7/2}$ shell. For comparison, results from
  Brown {\it et al.} \cite{Brown02}
  and from the Coulomb prediction \cite{Bentley07} are also shown.}
  \label{fig2}
 \end{center}
\end{figure*}

In the GXPF1A calculation within the full $fp$ shell, the INC
interaction for the $f_{7/2}$ shell has terms (see Eq. (\ref{eq:8}))
$V_{pp}=\beta_{pp}V^{J=0}_{pp}$, $V_{nn}=\beta_{nn}V^{J=0}_{nn}$,
and $V_{pn}=\beta_{pn}V^{J=0}_{pn}$, where $V^{J=0}_{pp}$,
$V^{J=0}_{nn}$, and $V^{J=0}_{pn}$ are, respectively, the $pp$,
$nn$, and $pn$ pairing interactions for the matrix elements having a
unit value. The parameters $\beta_{pp}=-22.5$, $\beta_{nn}=77.5$,
and $\beta_{pn}=-55.0$ (all in keV) are chosen so as to reproduce
the experimental CDE and TDE data. Fig. \ref{fig2}(a) shows that the
calculated CDE with and without the INC interaction can describe the
experimental data reasonably well. However, we find that only with
the isovector and isotensor interactions can the calculation
correctly reproduce the observed reduction in staggering magnitude
of $\Delta{\rm CDE}/Z$ for the mass region $A=45-51$, as shown in
Fig. \ref{fig2}(b), and the large experimental TDE shown in Fig.
\ref{fig2}(c). Without the INC nuclear interaction, the calculated
staggering magnitudes for $\Delta{\rm CDE}/Z$ are clearly larger
than the data and the TDE values are close to the Coulomb
prediction, but smaller by about 150 keV than the experiment. In the
JUN45 calculation, the INC interaction is not included.

The underlying physics is that inclusion of the isovector force in
the $f_{7/2}$ shell modifies interactions between the nucleons.
Namely, $pp$ ($nn$) now becomes more attractive (less attractive),
which results in an increase (decrease) of the proton (neutron)
pairing gap. To see its influence on $\Delta{\rm CDE}$ directly, we
rewrite Eq. (\ref{eq:2}) as
\begin{equation}
 \Delta{\rm CDE}(A,T)
= 2(-1)^{Z_{>}} \left[\Delta_{\pi}(Z_{>},Z_{>}) -
\Delta_{\nu}(Z_{>},Z_{>})\right],
  \label{eq:3}
\end{equation}
in which $\Delta_{\nu}(Z_{>},Z_{>})$ and $\Delta_{\pi}(Z_{>},Z_{>})$
are the three-point odd-even mass differences for neutrons and
protons, respectively, which are regarded as measures of the
neutron- and proton-pairing gap \cite{Satula98}. The occurrence of
the odd-even staggering can then be explained by the differences
between proton- and neutron-pairing gaps, with the factor
$(-1)^{Z_{>}}$ originating from number parity. Without the isovector
force in the $f_{7/2}$ shell, calculations give an overly strong
staggering for $A=43-51$ (see Fig. \ref{fig2}(b)). Now with
inclusion of the isovector force, an increasing difference between
$\Delta_{\pi}$ and $\Delta_{\nu}$ is obtained. With $Z_{>}=$ odd
(even), the factor $(-1)^{Z_{>}}$ in Eq. (\ref{eq:3}) is negative
(positive) for $A=41, 45, \dots$ ($A=43, 47, \dots$). As compared to
the results without the isovector force, this obviously leads to a
decrease in $\Delta$CDE for the sequence with odd $Z_{>}$ and an
increase for even $Z_{>}$, thus reproducing the observed reduction
of staggering magnitudes shown in Fig. \ref{fig2}(b). On the other
hand, since inclusion of the isotensor force makes $pn$ more
attractive than the average of $pp$ and $nn$, the last term in Eq.
(\ref{eq:4}) becomes smaller, thus increasing the TDE for $A=42-54$,
as shown in Fig. \ref{fig2}(c). We note that the calculations do not
support the apparent change in the staggering phase at $A=69$ in the
experimental $\Delta{\rm CDE}$. This may suggest \cite{Tu13} that
the mass of $^{69}$Br \cite{Rogers11} was measured for an isomer,
not for the ground state. From the present calculations, we find
that $nn$ is $\sim$0.8\% more attractive than $pp$, and $pn$ is
$\sim$2.5\% stronger than the average of $nn$ and $pp$. These ratios
are in accord with those estimated from the nucleon-nucleon
scattering data \cite{Machleidt01a}.

For the heavier mass region with $A=55-67$, the calculated CDE
differences are also in a good agreement with the observed large
staggering (see Fig. \ref{fig2}(b)). The sudden drop in TDE at
$A=58$ (Fig. \ref{fig2}(c)) and the corresponding drop in
$\Delta$TDE at $A=54$ (Fig. \ref{fig2}(d)) are correctly reproduced.
For nuclei below $A=54$, since nucleons occupy mainly the $f_{7/2}$
shell, the added INC interaction shows a significant effect, which
correctly describes the observed large TDE, as discussed above. For
the triplet nuclei with $A=58$, however, two nucleons occupy the
$p_{3/2}$ orbit and do not feel an INC interaction, and therefore,
the TDE decreases drastically. Thus, in our calculation the observed
sudden drop in TDE at $A=58$ may suggest that the INC nuclear
interaction is less important for the normal-parity $p_{3/2}$ and
$f_{5/2}$ orbits. Differences between the GXPF1A and JUN45
calculations are found above $A=$ 69 in Fig. 2, which are attributed
to the contribution from the $g_{9/2}$ orbit.

\begin{figure*}[t]
\centerline{
\mbox{
\includegraphics[width=6.8in]{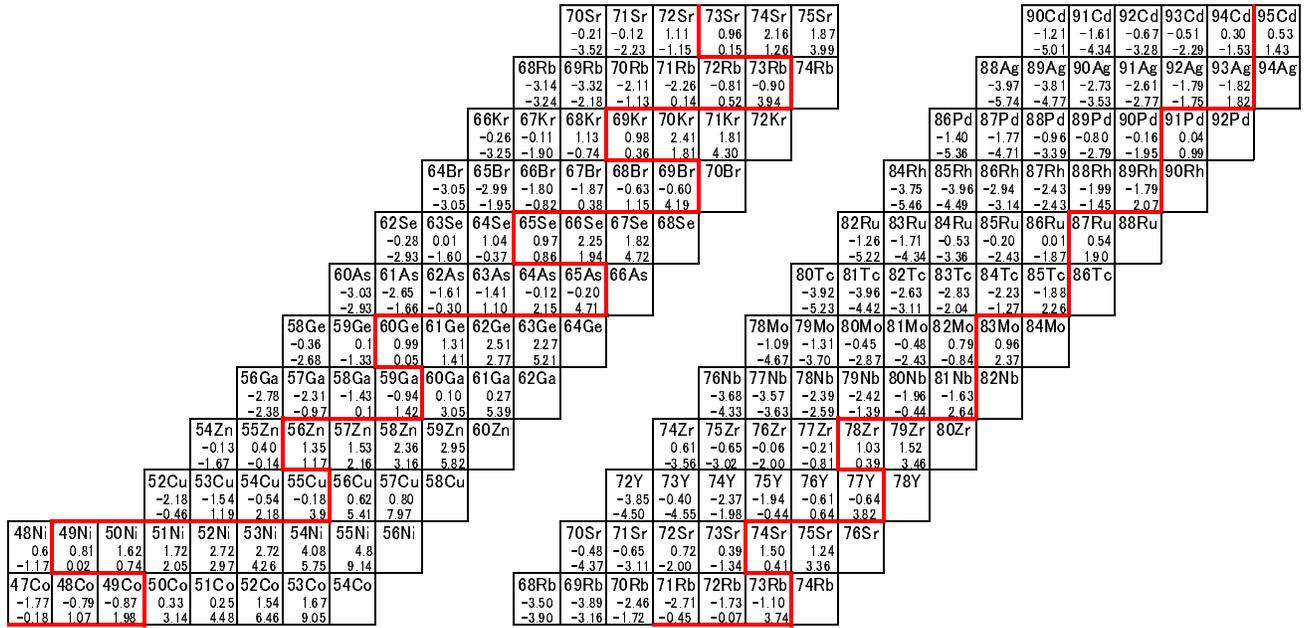}}}
\caption{(Color online) Calculated one- and two-proton separation
energies for odd-mass nuclei with isospin $T=1/2,3/2,5/2$ and for
even-mass nuclei with $T=1,2,3$ using the GXPF1A (left) and the
JUN45 (right) interaction. In each box, the first number denotes
one-proton separation energy and the second denotes two-proton
separation energy. Thick (red) lines indicate the proton drip-line.
} \label{fig3}
\end{figure*}

On the basis of the successful CDE calculation as presented in Fig.
\ref{fig2}, now we try to map the proton drip-line by evaluating
one- and two-proton separation energies. According to Eq.
(\ref{eq:1}), we use the shell-model CDE and the observed binding
energy $BE(T,{T_{z}}_{>})$ for the nucleus from the neutron-rich
side to predict the binding energy $BE(T,{T_{z}}_{<})$ for the
proton-rich analogue nucleus. For binding energies above $A=$ 82
where no data are available, we simply adopt the Audi-Wapstra
extrapolation from AME'03 \cite{Audi03}. The agreement between the
calculated and experimental binding energies is very good within an
rms deviation of about 100 keV. Figure \ref{fig3} shows the
calculated one- and two-proton separation energies denoted in each
box by the first and second numbers, respectively. The thick (and red)
lines represent the proton drip-line beyond which the one-proton and/or
two-proton separation energies become negative. The INC term for the
$f_{7/2}$ shell is included in the calculation with the GXPF1A
interaction. The existing data for $^{60}$Ga \cite{Blank95}, $^{64,65}$As
\cite{Robertson90,Tu11}, and $^{69}$Br \cite{Rogers11,Schury07}
indicate that these nuclei are unbound. The experimental separation
energies of $^{63}$Ge, $^{67}$Se, and $^{71}$Kr \cite{Tu11} suggest
that they are bound. In the graph on the right, the experiments
indicate that $^{77}$Y and $^{82}$Mo are bound while no evidence was
found for $^{81}$Nb and $^{85}$Tc. As one can see, most of our
results are consistent with the current experimental information.
Figure \ref{fig3} also suggests several candidates for proton
emitters.

In summary, we have investigated effects of the isospin
nonconserving forces that cause characteristic shell changes near
the $N=Z$ line. Large-scale shell-model calculations were performed
by employing two modern effective interactions (GXPF1A and JUN45)
for the corresponding mass regions with inclusion of the Coulomb
plus INC nuclear interactions. We concluded that the INC forces are
important for the $f_{7/2}$-shell nuclei, but not for the upper
$fp$-shell. This conclusion is consistent with those found in our
previous papers \cite{Kaneko10,Kaneko12}. No conclusion about the
INC forces can currently be drawn for heavier nuclei with $A=70-95$.
Consequently, we calculated one- and two-proton separation energies
to map the proton drip-line. Our calculation provides many new
predictions for the $fpg$ shell region up to $A=95$, which may be
relevant to the discussion of the rp-process of nucleosynthesis
\cite{Schatz01}. The results shown in the present Letter should be
tested by future experiments on proton-rich nuclei of the heavy mass
region.

Research at SJTU was supported by the National Natural Science
Foundation of China (Nos. 11135005 and 11075103) and by the 973
Program of China (No. 2013CB834401).



\end{document}